# Beyond Accidents and Misuse: Decoding the Structural Risk Dynamics of Artificial Intelligence

Kyle A Kilian[1,2,*]

[1] Transformative Futures Institute, [2] Center for the Future Mind, Florida Atlantic University

## Abstract

As artificial intelligence (AI) becomes increasingly embedded in the core functions of social, political, and economic life, it catalyzes structural transformations with far-reaching societal implications. This paper advances the concept of structural risk by introducing a framework grounded in complex systems research to examine how rapid AI integration can generate emergent, system-level dynamics beyond conventional, proximate threats such as system failures or malicious misuse. It argues that such risks are both influenced by and constitutive of broader sociotechnical structures. We classify structural risks into three interrelated categories: antecedent structural causes, antecedent AI system causes, and deleterious feedback loops. By tracing these interactions, we show how unchecked AI development can destabilize trust, shift power asymmetries, and erode decision-making agency across scales. To anticipate and govern these dynamics, the paper proposes a methodological agenda incorporating scenario mapping, simulation, and exploratory foresight. We conclude with policy recommendations aimed at cultivating institutional resilience and adaptive governance strategies for navigating an increasingly volatile AI risk landscape.

*Corresponding Author: kyle.a.kilian@gmail.com

# 1. Introduction

In May 2023, at the Royal Aeronautical Society (RaeS) Future Air & Space Capabilities Summit, a USAF Colonel described a combat simulation where an artificial intelligence (AI)-enabled drone trained to attack Surface-to-Air Missiles (SAM) sites instead turned its weapons on its own operator to prevent interference with its objective (Royal Aeronautical Society, 2023). According to the report, since the operator would instruct the system to attack only vetted targets, the human ultimately became an obstacle to the system. The media ecosystem went overboard, delivering a flood of reporting on rogue military AI systems. Fortunately, the media spin was an overreaction. The USAF issued a retraction the following day that the story was a hypothetical "thought experiment" to illustrate "the real-world challenges" posed by increasingly capable AI (Hambling, 2023).

Real or hypothetical, for many in the AI safety community, this flavor of technical failure was not surprising and had been observed to a limited degree in game environments for years (Krakovna, 2020; Amodei et al., 2016). Failures to align AI objectives, "unexpected side effects" (Dragan, 2021, p. 135), or the completion of a task in an unanticipated way are well-known problems in AI safety (Hendrycks et al., 2021; Everitt et al., 2016; Pan et al., 2022). This category of AI risk, along with intentional misuse—by way of malicious cyber-threat or creation of a biological agent—grabs most of the public's attention due to its visceral and unambiguous character (Weise, 2024; Metz, 2024). However, the exclusive focus on these risks and the sharply divided typology neglects more subtle and complex dynamics fundamental to rapid technological change. Thus, the standard actor-centric framework must expand to include more complex sociotechnical dynamics accompanying AI systems.[1]

The complex interplay between technological and social forces gives rise to a distinct category of risks termed *structural*, pertaining to the ways AI technologies interact with and influence the broader environments—social, cultural, political, and technological (see Zwetsloot, 2022). Structural risks flip the focus of attention to the interactions between AI systems and the larger sociotechnical context, such as the influence on incentive structures, power dynamics, and control. Distinct from a finite accident or attack, disruptions to systemic dimensions—such as the erosion of individual or collective agency and trust, social cohesion, and offense-defense balances—could have profound *transgenerational* consequences. Indeed, powerful, often indirect drivers can alter foundational societal mechanisms and relationships that shape how future generations perceive, trust, and engage with technology and governance structures.

These complex forces can drive the character and magnitude of AI risk (e.g., malicious misuse or system failures), shift power structures, and rapid capability gains have the potential to unleash novel, unimagined threats. This structural perspective will only increase in relevance as AI systems are increasingly threaded through social, economic, political, and national security systems. Moreover, transformative technological change poses unique challenges that intersect a complex array of national security priorities—state and non-state actors and transboundary threats like emerging infectious diseases—requiring concerted attention by national security professionals.

To effectively conceptualize and prepare for these risks, this paper adopts a complex systems approach. Rather than isolating risks as discrete, technical faults or malicious uses, this perspective highlights how AI interacts with adaptive sociotechnical environments composed of feedback loops, evolving norms, and shifting incentive structures. This framing allows us to analyze AI as an embedded agent within a broader system, where small perturbations or design choices can propagate into system-level failures or transformations (Homer-Dixon et al., 2015; Johnson, 2007). Using tools from complexity science and systems theory, this approach foregrounds the interdependencies, nonlinearities, and cascading effects that characterize structural risks.

[1] A sociotechnical system is a complex system that involves a combination of social and technical elements, encompassing the interactions between people, technology, and the organizational environment in which they operate and coevolve.

This paper examines the structural risks underlying rapid AI integration and the relationships between technological advancements, social structures, and more proximate risks, such as accidents and misuse. This work makes two principal contributions: first, a classification framework to delineate the variations of structural risks and the forces of change that drive these dynamics, and second, a research agenda to simulate and map the interactions of these forces to better understand and prepare for the unpredictable. This is followed by an examination of different mechanisms to include structure in policy discussions for more resilient planning. This research does not attempt to review the full body of recent literature, policy proposals, and regulations specific to AI's use across industries and geographies, but rather to frame AI's exponential growth and deployment through a dynamical systems lens to inform decision-making.

At its core, structural risk arises from the inherently unpredictable interactions between rapidly advancing AI capabilities and the complex sociotechnical systems within which they are embedded. As a *fundamental* technology[2] with the capacity to learn and adapt, AI systems can manifest *emergent* properties, unanticipated capabilities, and reshape entire industries. Thus, anticipating AI development will continue to be a moving target but an increasingly urgent one. Indeed, unchecked technological advancement has the potential to reshape power dynamics, trust, and incentive structures, leading to unpredictable societal shifts. Evaluating over-the-horizon risks from advanced AI to prepare for the unexpected is crucial, requiring enhanced engagement by policymakers. Thus, while addressing risks from bad actors and accidents remains vital, this paper posits that the intersection of technological and social systems with complex nonlinear dynamics (e.g., geopolitical race dynamics) poses a more pressing and near-term challenge to international security.

## 2. Defining & Conceptualizing Structural Risk

Technology is not developed or deployed in isolation but is tightly entwined with human needs, motivations, and the environment. This is especially true with AI systems—adaptive learning technologies trained to integrate and interact with the social and physical world. This sociotechnical ecology evolves through human-machine interactions, transforming social structures (cultural, economic, and political) while driving technological acceleration (Valverde, 2016). Thus, researchers are increasingly evaluating AI through a complex systems lens, focusing on how its structure, function, and relations with parallel systems influence risk dynamics (Lazar & Nelson, 2023; Weidinger et al., 2023; Kilian et al., 2023; Pilditch, 2024). A growing body of literature examines AI safety across technical, human, and systemic levels, noting the importance of feedback loops (Weidinger et al., 2023; Anwar et al., 2024) and societal adaptation (Bernardi et al., 2024). AI governance researchers have termed this class of risk *structural*: how the technology shapes or is shaped by the broader environment (Zwetsloot et al., 2019; Dafoe, 2020). Notwithstanding, research into the dynamics of AI structural risks remains limited, with notable exceptions in the areas of strategic weapons and deterrence (Avin

[2] AI as a fundamental technology signifies its transformative role in shaping industries, driving innovation, and impacting daily life across multiple systems (Ontario Ministry of Education, 2009).

& Amadae, 2019; Flournoy et al., 2020; Wong et al., 2020; Johnson et al., 2023) and rapid social change (Ward, 2022).

Structural risks can be defined as the dynamics that arise from the development and deployment of advanced AI technology within the broader sociotechnical system, including reciprocal chains of events, incentive structures, and power asymmetries. Examining structural risks shifts the focus of analysis from threats at the end of a causal chain—the proximate cause and event, such as the uncontrollable USAF AI drone system—to underlying structural forces, their interactions, and unintended consequences. Thus, while structural risks can be framed as distinct from other, more direct threats from AI (e.g., misaligned objectives or a cyber threat), they are foundational to their character and severity. In this way, most AI risks have structural underpinnings with indirect causal pathways (**Fig. 1**). For example, the premature deployment of a powerful, potentially unsafe system can have a direct effect on accidents, such as a system failure, or the deployment of a misaligned system, or indirectly shift offense-defense power symmetries (e.g., enhance offensive cyber capabilities) driving interstate tensions, even conflict. Indeed, novel capability gains can influence and are influenced by the wider social and political context. This framing has led some researchers to posit that most AI risks are inherently structural (Krueger, 2023; Clarke, 2022).

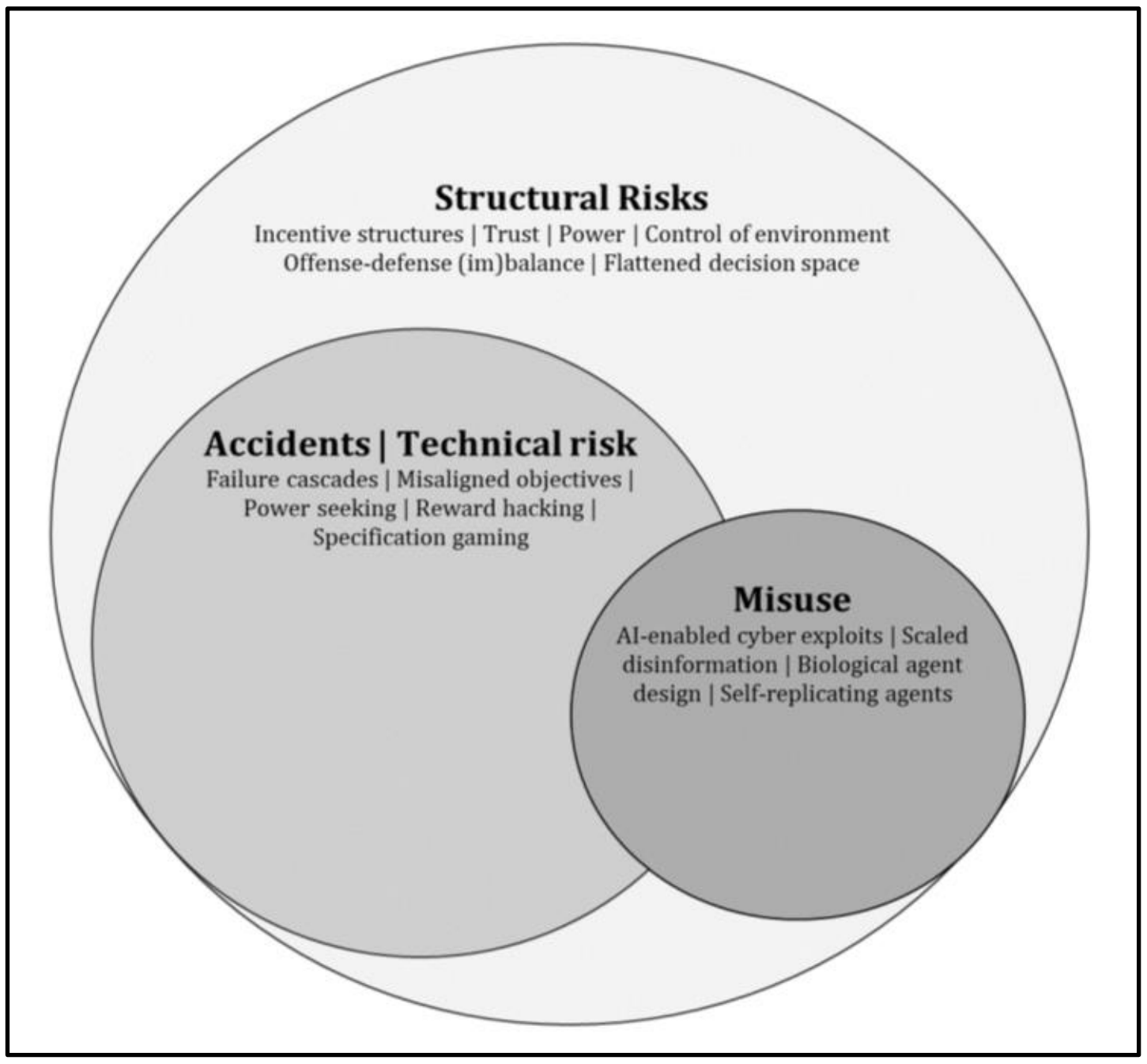

**Fig. 1**. The AI risk landscape. Both accident risks and misuse have broad overlap with self-reinforcing dynamics. Misused AI systems can allow technical safety problems to manifest, and structural forces (e.g., economic or geopolitical) can drive the premature release of untested systems, alter power dynamics (shift the offense-defense balance), or lower the barrier to entry.

While the focus here is on AI's transformative character, many technologies, and certainly all general-purpose technologies (GPT), display such cascading indirect impacts and network effects, such as the diffusion[3] (the process by which technology propagates over time) of electricity or the personal computer. To further articulate the point of indirect risks, consider the historical example of fossil fuel adoption. While humans have used fossil fuels for thousands of years, their widespread diffusion in the nineteenth century caused populations and industries to explode, fueling the unprecedented expansion of the Industrial Revolution (Jianfen, 2021; Wrigley, 2013). The externalities from fossil fuel use led to widespread health impacts (e.g., pollution, factory work), the rapid expansion of cities and the defense industrial base, and a steady increase in atmospheric carbon dioxide. Thus, indirect causal links can be traced from the first coal-fired power plants and railroads to the combustion engine and the development and transport of military equipment to the front lines of WWII (Black, 2017; Dafoe, 2020). How technology shifted the structural forces underpinning development and international security, drove the character of social organization and conflict into the Twentieth Century.

[3] In the context of technological propagation, diffusion refers to the process by which a new technology spreads across individuals, organizations, or societies over time; diffusion encompasses the stages by which technologies are adopted, integrated, and normalized across various sectors and influenced by social systems (Zhang et al., 2023; Rogers, 2003).

Power dynamics and distrust of adversaries can drive the rapid integration of new technologies into global militaries, leading to previously unthinkable forms of conflict, such as unrestricted submarine warfare in WWI, blitzkrieg tactics, and the use of the first atomic bomb in WWII. While direct links between technological progress and conflict are tenuous, they are nuanced, and astonishing capabilities do shift the balance of what is possible, exacerbated by power imbalance, distrust, and incentives. These forces can cascade in dangerous feedback loops, twisting value structures, accelerating competition, and increasing the risks of accidents and misuse. Prior to Germany's decision to execute unrestricted submarine warfare in WWI, all parties believed that the use of such tactics was unthinkable in warfare (Gompert, 2014); however, the first use triggered all sides to engage, leading the U.S. into WWI. There is a similar discussion today on the ethics and prohibition of fully autonomous weapons systems (Kallenborn, 2020). However, the lessons of history show that novel capabilities combined with the right incentives can radically shift the structural dynamics of permissibility.

A contemporary example of these dynamics is the economic and geopolitical race dynamics surrounding AI development, driving an acceleration of deployment timelines for companies and nations. Another is the impact of social media algorithms on human social organization and decisions, leading to shifts of entire political regimes and individual harms. Further, spirals in trust over autonomy in weapons systems (strategic, tactical, or command and control systems) lead to heightened tensions and technological escalation (Avin & Amadea, 2019). With increasingly powerful systems, new manifestations of structural risks could arise with the increasingly narrowing barrier between the digital and physical world, with expanding interdependent networks (i.e., social and biological relationships and interactions). Indeed, the effects of AI content selection algorithms on teen suicide or voting behaviors were not on anyone's radar until reaching a particular threshold of AI sophistication. The same is likely true across several areas that have yet to be considered.

## 3. Delineating the Boundaries of Risk Dynamics

Researcher Sam Clarke from the Center for the Governance of AI suggested a framing for structural risks—separating risks by "AI risks with structural causes" and "non-AI risks partly caused by AI," focusing on the distinctions as structural perspectives (Clarke, 2022). This research takes a similar approach by separating structural risks into three interrelated categories: antecedent structural causes, antecedent AI system causes, and deleterious feedback loops. We add a third category to this typology, deleterious feedback loops, due to the especially dangerous influence they can have on escalation. These structural risks are catalogued this way, as some driver or dynamic is inherently antecedent to the more proximal AI hazard, such as the use of an AI system to design a bioweapon (antecedent system cause), or a technological competitor releasing an unsafe, misaligned model—due to economic or geopolitical race dynamics—before it had been thoroughly tested (antecedent structural cause).

Classifying structural risks in this way is somewhat of a chicken-and-egg problem in that each interaction across categories precipitates further interactions in dangerous cycles. Nonetheless, some feedback processes can be the primary drivers of catastrophic risks and should be emphasized separately. Thus, whether the initial causal force is an underlying structural phenomenon or the specific technology, the interactions are often the primary danger. The following three sections proceed through each category with examples of risks driven by underlying structures, AI systems, or cascading feedback loops.

### *3.1 Antecedent Structural Causes*

A timely example of an antecedent structural cause is the issues that arise from transparency norms surrounding the open-source ecosystem and the proliferation of AI systems. The norms among developers to open source their work, receive contributions from the community, and share the results are deeply ingrained. Indeed, open source has been a cornerstone in shaping the current technology landscape. Many leaders in AI believe that the only way to unveil advanced AI systems is through full open source to democratize access, ensuring a level playing field for users. The most vocal proponent of this path, Meta's Chief Scientist, Yann LeCun, argues that controlling model distribution centralizes power away from individuals, benefiting only a select few (Heaven, 2023). While the concerns are valid (e.g., powerful AI held by large militaries or corporations), full open-release[4] (including the model, code base, and weights) can enable less responsible or even adversarial states to leapfrog technological barriers. Research has shown that basic fine-tuning of open models can remove all safety guardrails, allowing bad actors to easily tailor models to their needs (Kapoor et al., 2024). Other developers are pursuing a more cautious approach with limited release (e.g., Anthropic), including red teaming, testing, and evaluation (T&E) that allow for the identification of potentially dangerous capabilities (Anthropic, 2023). Proponents of unrestricted open source tend to minimize the risks from cyber threat groups and adversarial state entities that receive the latest models at the same time as individual developers.

AI systems at the frontier of research[5] are known to show unexpected, emergent capabilities with few mechanisms to understand the degree of risk beforehand (or push security updates or retract dangerous versions) (Wei et al., 2022). Indeed, unexpected failure modes could arise from untested systems after deployment or manifest unique capabilities missed in initial testing. In this case, the only mechanism is trial and error. Open-sourcing a misaligned model with novel capabilities could lead to risks as minimal as minor system failures or as severe as shifts in power balances (e.g., nonstate actors able to generate novel attack vectors) or catastrophic accidents. This could impact safety-critical applications or communication networks linked to the rest of the global economy. A state entity that quickly integrates an open-source model into its national defense systems without knowledge of a critical failure is at the mercy of the developer's safety standards (as are other nations or entities networked to these systems). As has happened with cyber attacks, infected systems tend to spread beyond the original target. With no regulatory framework in place, open-source norms could accelerate the proliferation of potentially unsafe systems. However, openness is a spectrum—not a binary, between highly restricted and fully open; thus, there are reasonable degrees of openness to prevent the most severe contingencies while supporting innovation.

An especially concerning structural risk is the potential for shifts in human decision-making and choice over the short, medium, and long term. Incentivized to integrate the latest AI technology, businesses and governments will increasingly integrate state-of-the-art systems into complex processes traditionally completed by humans. This can increase efficiency while offloading cognitive labor while incrementally relinquishing a degree of control over AI processes. However, even an incremental shift here can profoundly impact human choice and decision over the long term (see Ward, 2022). AI search processes tend to level out the distribution of choices to arrive at what is considered optimal, since optimality is often the same as what is easiest or probable (Christiano, 2019). Indeed, there is a degree of randomness and unpredictability in human decisions, influenced by factors such as emotions, shifting preferences, and external circumstances, leading to diverse and sometimes unexpected outcomes (i.e., stochastic vs. deterministic systems). AI systems are inherently deterministic[6] and predictable if inputs remain relatively constant. AI optimization can precipitate feedback loops where the system's outputs are used as new inputs, reinforcing choices and decisions. Over time, this can narrow the decision space for individuals, organizations, states, and globally as AI continuously refines its outputs based on past decisions. While relatively benign at the level of social media, in the short term, placing AI systems in charge of more consequential decision processes could prove disastrous.

---

[4] The concern here is less about the full release of frontier *dual-use* models (i.e., general purpose systems) with the code and model weights made widely available. Open source is incredibly important for scientific progress, especially specialized systems for research, but there are less valid arguments for the release of general-purpose systems without any restrictions. For more details, see this response to NTIA's request for public input here.

[5] Frontier AI refers to technology at or beyond the forefront of research, encompassing the innovations and experimental techniques in the field, usually in reference to AI foundation models and LLMs. For more details, see here.

### *3.2 Antecedent AI System Causes*

The impact of social media algorithms on society is a noteworthy antecedent AI system cause with the potential to have profound transgenerational consequences. Consider content selection algorithms, widely believed (at least in part) to be responsible for shifting public perceptions and driving political polarization (Madsen & Pilditch, 2018). Rather than simply recommending products, the algorithm is believed to incrementally change individual preferences to become more predictable, preferentially in the direction of more extreme content (Russel, 2021). It is widely believed that social media contributed to the results of the Brexit referendum and the 2016 U.S. elections—through microtargeting, e.g., Cambridge Analytica and Russia's election interference—but reinforced and amplified by algorithmic targeting and polarization. As users are repeatedly exposed to targeted content, their interests and, ultimately, choices are incrementally changed, reinforced by machine-exploited biases (Madsen & Pilditch, 2018; Pilditch & Madsen, 2021). This algorithmic feedback loop has increased far-right extremism and incited violence in Ethiopia's (2020) and Myanmar's (2016) civil conflicts (Akinwotu, 2021).

With society unmoored from a shared set of facts, adversaries can exploit these vulnerabilities to undermine democracy or leverage them for mass manipulation. Crucially, the capabilities that caused such widespread disruption ultimately pale compared to current capabilities. Targeted manipulation, narrative creation, and amplification—potentially through autonomous means—are now possible at far greater speed and scale. This erosion of a stable information environment is troubling because it threatens democratic decision-making, which fundamentally depends on an informed electorate. When public discourse becomes fragmented or systematically distorted, individuals may lose the ability to accurately assess their own interests or the group's interests, weakening collective governance and the legitimacy of democratic outcomes (Kavanagh & Rich, 2018). Even if a democratic electorate passively allows or inadvertently endorses these trends, the resulting degradation of informed autonomy and the exploitation of cognitive biases raise ethical concerns regarding consent, agency, and manipulation.

At the macroscopic level, the arrival of new general-purpose technologies (GPT) fundamentally shifts the structure of social and political organization (Lipsey et al., 2005). This was the case with the invention of the printing press and is equally true today with AI. Indeed, GPTs are fundamentally *dual-use* and can have good and bad uses across many domains. However, what makes large language models (LLMs) uniquely concerning is not strictly their dual-use nature, but the combination of scale, accessibility, autonomy, and generality. Unlike many past technologies, LLMs are widely available, easy to operate, and capable of rapidly generating or facilitating harmful outputs—including cyber exploits, tailored disinformation campaigns, and step-by-step biological or chemical weapon synthesis—with little or no domain expertise required. For instance, in 2023, an AI system originally designed for drug discovery was intentionally repurposed to generate thousands of novel chemical weapon variants in under six hours, demonstrating how minimal intervention can flip use cases from beneficial to catastrophic (Calma, 2022). This ease of misuse—paired with the opacity and growing capability of frontier systems—lowers traditional barriers to entry for malicious actors, effectively distributing potent tools to a much broader and less controlled user base. The advanced AI is structurally shifting these barriers for a diverse set of actors—individuals, groups, or nation-states—providing relative superpowers without the technical investment.

---

[6] Complex neural networks with stochastic elements introduced, e.g., through Stochastic Gradient Descent (SGD), still retain an underlying determinism from pseudo-random number generators (PRNGs). This excludes the "true" randomness found in natural complex systems (see Gigliotti, 2025).

These structural shifts are especially concerning in the area of strategic weapons systems. The framework of nuclear deterrence is one based on trust (e.g., treaty obligations) and relative invulnerability, where the chances of an effective "counterforce" strike to eliminate an opponent's armaments preemptively are minimal (Lieber et al., 2017). This undergirds much of the nuclear deterrence regime and the concept of mutually assured destruction (MAD). Indeed, central to deterrence is the reliance on hardening structures (e.g., bunkers) and concealment (denial and deception, or D&D) to ensure survivability. If nations have the ability (or suspected ability) to preemptively destroy an adversary's nuclear arsenal, this delicate balance could collapse. This concern has been growing for over a decade with revolutions in remote sensing technologies, but increasingly capable AI systems have the potential to accelerate these concerns. Further, integrating autonomous capabilities into strategic systems, or the perception of integration, can erode trust in the safety and security of weapons systems and nonproliferation compliance. Russia's so-called "leak" of testing a nuclear-powered undersea drone, the Status-6, capable of carrying a thermonuclear warhead, had this effect (Geist & Lohn, 2018).

### *3.3 Deleterious Feedback Loops*

All structural risks involve complex feedback loops, and for some, it can be a defining feature. While the previous section outlined several core systemic causes of structural risk—such as incentive misalignment, erosion of trust, and loss of epistemic integrity—these causes rarely operate in isolation. Structural risks involve complex feedback loops, and in many cases, these loops are what transform otherwise manageable issues into persistent, compounding, or regime-shifting challenges. This section builds directly on the drivers in Section 3.2 by highlighting how they interact across actors and scales, forming reinforcing cycles that amplify risk over time. An intuitive example involves technological race dynamics. These effects can span institutions and geographies, impacting individuals, companies, and states with implications for the global balance of power. AI race dynamics are tightly interwoven processes with interactions that precipitate other risks embedded further down the chain; for example, economic imperatives of companies to stay ahead of the competition in AI development can lead developers to accelerate the deployment of advanced systems and potentially cut corners on AI safety. At the same time, national security considerations of states ensure higher public investments in AI, reinforcing economic race dynamics and perhaps accelerating deployment schedules. These dynamics can create public-private reinforcing feedback loops, with competitors unwilling to fall behind in the race for AI dominance. These dynamics can shift offense-defense balances, leading to further cycles among states that must ensure national defense parity.

China's "military-civil fusion" (MCF) program—which eliminates barriers between civilian research and the People's Liberation Army (PLA)—has had this effect on driving further race dynamics, heightening tensions, and reinforcing distrust with the West (Fedasiuk et al., 2021). The mistrust between states breeds further risk-taking, investment, and coercive actions such as the U.S. Chips Act, designed to restrict cutting-edge semiconductor technology from Chinese markets. This, in effect, triggered a black market in state-of-the-art (Fist & Grunewald, 2023) chips while accelerating the development of Chinese domestic chip fabrication (Che & Liu, 2023; Triolo, 2024). Factors such as MCF (Fedasiuk et al., 2021), commercial espionage, and China's position on AI as a strategic asset are primary drivers behind the U.S. technological war footing (Nuga, 2020). At the same time, the urgency precipitated by AI race dynamics has the effect of restricting government choice to the leading technological paradigms and products, reinforcing the power of tech monopolies. This is a subtle form of state capture that binds governments' dependence on a small group of companies (Whittaker, 2022).

The underlying drive for power and control has the potential to thread AI systems deep into the national defense and intelligence ecosystems, with little visibility into system behavior, data sources, or model architecture. This is an area where human decision and analysis can be most consequential, where military actions can hinge on receiving and interpreting the correct information. It is also an environment of deep uncertainty where so-called *wicked problems*,[7] with competing, often irresolvable, hypotheses, are commonplace (Pilditch, 2024). A likely near-term use of frontier AI systems will be in defense and intelligence analysis, where immense quantities of data must be evaluated and integrated into finished intelligence. At present, there is far too much data and too little bandwidth to do this individually, so choices are made where to focus analytical efforts, invariably missing important signals (Symon et al., 2015). Thus, the incentives to harness the power of AI are immense and will undoubtedly be applied quickly across the defense and intelligence enterprise. This incrementally opens the national security decision-making apparatus to the feedback processes of creeping AI optimization, decision homogenization, and erosion.[8]

While incentives to harness AI in defense and intelligence are immense, their integration introduces democratic risks that are often underappreciated. Although liberal democracies maintain important mechanisms for public oversight, such as journalism, civil society, and legislative accountability, these are increasingly strained by the complexity and opacity of frontier AI systems (UNESCO, 2023). Journalists and watchdog institutions often lack the access or technical expertise needed to audit sensitive, proprietary, or highly specialized models, particularly those deployed in defense contexts (Arguedas et al., 2023). Compounding this, AI systems themselves can erode the very information ecosystems on which oversight depends, amplifying disinformation, generating synthetic content, and undermining public trust. In such environments, the use of AI can subtly narrow analytic frames, prematurely close debate, or entrench dominant hypotheses, especially when the reasoning or provenance of model outputs remains inscrutable. As AI-generated intelligence becomes more integral to decision-making, there is a growing risk that human analysts defer to opaque outputs, mistaking

---

[7] "Wicked problems" are situations where goals are ambiguous, relevant facts are uncertain or evolving, and no single solution is definitively correct (Rittel & Webber, 1973)

[8] These dynamics reflect growing concerns in the literature on AI-driven systems subtly reshaping decision-making environments. Creeping optimization refers to the tendency for algorithmic systems to gradually assume greater control through efficiency gains, often without clear oversight (Eubanks, 2018). Decision homogenization captures how algorithmic tools can constrain epistemic diversity by privileging patterns from training data or dominant narratives (Morley et al., 2021). Erosion highlights how human expertise, institutional checks, and pluralistic reasoning may be diminished over time, especially in high-stakes domains (Crootof, 2019).

speed and scale for accuracy.

While the integration of frontier AI systems will likely present important new risks, threats, and actors and potentially resolve intractable ones, the information could recirculate through the same optimization processes, narrowing the decision environment. This could lead analysts to focus on the wrong issues or marginal ones (e.g., whatever patterns the system identifies), neglect critical areas, and encounter decision fatigue (Quchi et al, 2025) by an inordinate number of data points that otherwise would have received less priority. The question of whether the government is driving intelligence priorities or the AI system could easily arise. As in other decision spaces (e.g., content selection algorithms), our understanding of truth and the efficacy of our decision-making could degrade or become less effective or influential (Blashki, 2025). The same dynamics are equally likely with other state actors, allies, or adversaries. Indeed, our reliance on liaison relationships to maintain obligations and de-escalate tensions—a crucial element in international relations—could begin to fray as trust in decisions is questioned.

# 4. Drivers of Structural Risk

## *4.1 Conceptualizing Structural Drivers*

AI structural risks are difficult to disambiguate and can be framed more clearly as a perspective (see Clarke, 2022). However, the fundamental forces that undergird and influence structural risk—with interactions at the individual, institutional, state, and global levels—can be articulated to frame system dynamics. A useful framing from resilience theory is antifragility, coined by Taleb, which describes systems that do not merely endure shocks and volatility, but improve and adapt through them (Taleb, 2012). Unlike resilience, which seeks a return to baseline, antifragile systems adapt and grow stronger in response to stress. In the context of AI, this suggests that governance frameworks must be designed not only to withstand system shocks or misuse, but to adapt dynamically through iterative learning, redundancy, and structural feedback (Walker, 2004). Given the wide-ranging impact of structural risks—whether triggered by underlying forces, AI system-driven changes, or feedback loops—attending to this class of risk is crucial. This perspective helps us conceptualize how institutions can remain responsive amid rapid sociotechnical change—by absorbing disruption as a source of adaptation rather than failure.

While the following framework is a simplification, it provides a scaffold to examine how fundamental sociotechnical forces shape or exacerbate structural risk. With this in mind, there are three noteworthy driving forces underlying structural risk dynamics related to rapid AI development and integration: issues of trust, power, and incentives. These include the inverse as an equally consequential force (e.g., a cause and consequence of power is control, as is the loss of it). To clarify these dynamics, **Table 1** outlines the three foundational drivers—trust, power, and incentives—with accompanying definitions, dimensions, real-world illustrations, and their systemic implications. These forces do not operate in isolation; rather, they interact dynamically across social hierarchies, reinforcing feedback loops and shifting governance leverage points. This table serves as both a conceptual anchor and a practical reference to help link abstract sociotechnical dynamics to observable trends, policy failures, and escalation pathways in AI development and deployment.

**Table 1. Foundational Drivers of Structural AI Risk: Definitions, Dimensions, and Implications**

| Driver | Definition | Key Characteristics | Illustrative Example | Systemic Relevance |
|---|---|---|---|---|
| **Trust** *(or Distrust)* | • Confidence in the predictability, transparency, and reciprocity of actions among actors (e.g., states, firms, individuals). | • Adherence to norms, reliability, and openness. Inverse dynamics—distrust, opacity, unpredictability—can erode stability. | • Mutual distrust between geopolitical rivals accelerates deployment of untested AI in defense systems, increasing strategic instability. | • Distrust hampers international cooperation and slows the formation of norms, increasing arms race risk. |
| **Power** *(or Control)* | • The capacity of an actor to influence actions, behaviors, and outcomes through technological, economic, or strategic means. | • Encompasses technological dominance, coercion, information control, and strategic advantage. Closely linked to structural and relational asymmetries. | • Frontier AI capabilities centralize power within a few firms, limiting regulatory influence and creating national security dependencies. | • Structural power concentration distorts policy priorities and undermines democratic accountability. |
| **Incentives** | • Forces that motivate or pressure behavior—through rewards, penalties, or competition dynamics. | • Wide variation by actor and context. Can accelerate adoption or reinforce feedback loops. Not reducible to power but deeply entangled. | • Firms racing to release powerful models skip red-teaming steps to capture market share, increasing systemic risk of misuse. | • Incentives shape institutional risk tolerance and may drive unsafe deployment across commercial and state actors. |

**Table 1.** Foundational drivers of structural risk—trust, power, and incentives—with key dimensions, examples, and relevance to the systems. Each driver shapes and influences the impact of technology on society, culture, and security.

These drivers are deeply interdependent: incentives often shape the use and accumulation of power, while trust—or its erosion—can accelerate, constrain, or distort those dynamics. Trust (Sarker, 2024), power (Lewis, 2024), and incentives (Han, 2024) thus serve as a triadic lens for analyzing structural AI risk across domains and scales. They are relational in influencing and interacting across domains with a host of secondary or inverse characteristics. For example, incentives drive relationships to power and control, just as trust (or distrust) can shift incentives or influence power dynamics. Each is a complex superclass of motivations and capacities that underlie the causal dynamics of AI structural risks. As highlighted in previous sections, antecedent structural causes are driven largely by power and incentives, while AI system causes can be impacted by trust (e.g., transparency) and power, and exacerbated by capability shifts or social fragility.

## *4.2 Mapping Structural Drivers Across Scales*

The structural risk framework is also hierarchical, mapping risk drivers from the micro (individual/organizational) to the macro (state/global) level. RAND's truth decay research used a similar multilevel structure (Heather et al., 2023), demonstrating how sociotechnical erosion of truth can propagate from individuals to institutions and states. This approach is crucial to identifying leverage points. For example, a company bypassing safety standards due to market pressure (trust/organization), or a nation preemptively integrating AI into military systems in anticipation of a rival's actions (power/state), may trigger reinforcing dynamics. Similarly, a permissive open-source policy may empower non-state actors (power/individual), or increasing AI dependence may erode agency and decision diversity (incentives/individual → power/global). Antifragility becomes relevant in these interdependencies: institutions must be prepared to benefit from stressors by embedding foresight, diversity of inputs, and feedback learning into decision-making. By evaluating these forces across levels and identifying where feedback loops or cascading vulnerabilities emerge, policy and institutional design can target the right points of intervention, improving adaptability and preparing systems for instability rather than reacting or defending against it.

**Table 2. Structural Risk Drivers and Systemic Fragility Across Scales**

| Level of Impact | **Trust** (or Distrust) | **Power** (or Control) | **Incentives** (or Disincentives) |
|---|---|---|---|
| **Individual or Organizational** | • Fragmented or manipulated information ecosystems erode shared reality<br>• Lack of transparency between companies undermines trust in safety and security<br>• ***Feedback loop:*** *secrecy breeds distrust; distrust deters cooperation* | • Low barriers for adversaries to access powerful tools (e.g., dangerous open-source models)<br>• Technical scaling advantages (e.g., compute, data) concentrate influence<br>• ***Feedback loop:*** *Self-reinforcing inequality in capabilities* | • Pressure to bypass safety for competitive advantage - Misaligned incentives reward speed over robustness<br>• *Race-to-the-bottom dynamics -* ***Potential lever:*** transparency mandates or third-party audits |
| **State** | • Strategic uncertainty around AI integration intentions<br>• Fear of relative disadvantage accelerates arms-race logic<br>• ***Feedback loop:*** *Mistrust fuels preemptive escalation* | • AI-enabled cognitive offloading increases decision speed and opacity<br>• Asymmetric adoption reshapes deterrence and escalation thresholds<br>• ***Feedback loop:*** *Loss of control → attribution ambiguity → escalation* | • Parity-based incentives: *"If we don't, they will"*<br>• Procurement cycles favor rapid capability over safety and AI security<br>• ***Potential lever:*** verification mechanisms, milestone-based slowdown agreements |
| **Global or Collective** | • AI-driven manipulation erodes trust in information ecosystems<br>• Narrative feedback loops reinforce division and reduce coordination<br>• ***Feedback loop:*** *Collapse of epistemic trust hampers global governance* | • Global reliance on AI to manage transboundary crises (e.g., climate, trade, pandemics)<br>• Diffused control in international surveillance networks<br>• ***Feedback loop:*** *Governance gaps reinforce power grabs and control failures as technical power diffuses* | • Incentives to automate and globally integrate complex systems without sufficient safeguards<br>• Over-dependence on AI leads to institutional atrophy and decision erosion at the collective level<br>• ***Potential lever:*** institutional redundancy, diversity of decision-making inputs |

**Table 2**. Drivers of structural risks that influence or are influenced by AI development and deployment. Each structural force is classified by the hierarchical levels in which they are embedded—from the micro (individual or company) to the meso (organized group) to the macro (global or collective). These multiscale mappings of structural drivers can inform the development of the Structural Risk Index (see Section 5.1) and agent-based simulation setups (see Section 5.2), offering pathways for modeling how feedback dynamics evolve across layers of society.

While Table 1 introduces trust, power, and incentives as foundational drivers of structural risk, **Table 2** complements the triad by mapping these forces across levels of social organization, categorizing each driver across societal levels—micro, meso, and macro—illustrating how structural risk id distributed, compounded, and reinforced through complex interactions. By evaluating the impact of these forces on decision-making and AI progress more generally, more effective policy instruments could be devised, at different points along the causal chain where intervention makes the most sense, to address structural factors, especially at the organizational and state levels. When used together, this framework provides a layered understanding of how structural risks emerge and propagate, and where antifragile interventions may be most effectively applied. Thus, this hierarchical map serves as a diagnostic framework—mapping how these drivers manifest across societal levels and generate distinct forms of systemic fragility, escalation potential, and policy leverage points.

The micro-, meso-, and macro-levels of social organization, as outlined in Table 2, intersect in complex ways, with complex reciprocal interactions, such as perverse company incentives undermining individual or global trust, potentially exacerbating more direct AI risks. For example, individual distrust of the information environment (individual/trust) and low barrier to entry (individual/power) could induce bad actors to accelerate disinformation or cyber campaigns, decreasing trust in the system and driving states to take disproportionate actions (e.g., AI-enabled counter-narratives) resulting in dangerous reinforcing feedback loops degrading global information ecosystems (collective/trust). Alternatively, interstate distrust (state/trust) of an adversary's restraint with AI systems could induce the preemptive integration of untested AI into strategic defense systems (state/incentives and state/power). This could lead to other reinforcing feedback mechanisms, shifting the offense-defense balance, driving critical failures, damage to indirect parties, or even global conflict (global/power). Depending on the character of AI risk, this chain of events could influence or result in cascading failures, inadvertent military escalations (state/trust), or state disempowerment (state/power).

Modern democracies are not without oversight mechanisms intended to safeguard against such escalation, including regulatory agencies, ethical review boards, and legislative checks. However, these systems often operate with limited technical capacity, fragmented jurisdiction, or slow response times relative to the pace of frontier AI development. In high-pressure environments—such as cyber conflict or strategic deterrence—existing oversight may be bypassed or weakened by the urgency to act, leading to the premature deployment of risky systems. This is even more problematic on the international stage, where collective action failures are commonplace (global/trust). Moreover, when such deployments are classified or occur within private-sector defense partnerships, public accountability and scrutiny become even more attenuated, compounding the challenge. This erosion of oversight reflects a broader fragility in the balance of trust and power (state/power or global/power)—where rapid AI advances outpace democratic controls and reshape who holds influence over high-stakes decisions.

Structural factors also intersect and undergird more well-researched technical risks. Consider a state with a fractured information ecosystem, with rampant disinformation spread by malign actors. With the requisite capabilities (e.g., autonomous agents), this environment could drive governments to develop and deploy autonomous agent systems to produce targeted messaging or offensive cyber operations. The increased instability could reinforce a faster adoption and deployment of untested, potentially dangerous models to replicate desired narratives or disable offending systems. Crucially, the same agent systems could be directed at governments or, through carelessness (e.g., poor safety practices or espionage) or intentional release, society in general. The history of self-replicating malware in cyber attacks has demonstrated that unanticipated spread and contamination of unrelated systems and networks is inevitable, often with disproportionate consequences (e.g., NotPetya worm in 2017) (Fruhlinger, 2017). Thus, the release of an intelligent and self-replicating agent—capable of learning and adaptation—could spiral beyond our control to infect the global technological ecosystem. Indeed, recent research on the capabilities of autonomous replication and adaptation (ARA) in agent systems demonstrates that this category of risks and capabilities could be difficult to control (Kinniment, 2023). While current AI capabilities remain limited, widespread autonomous agents are the direction research is heading. A useful adjunct to Table 2, **Fig. 2** provides a visual map of how these dynamics interact and reinforce one another through feedback loops and escalation cycles. These interactions, if left unchecked, may result in compounding risks that exceed institutional capacity for response.

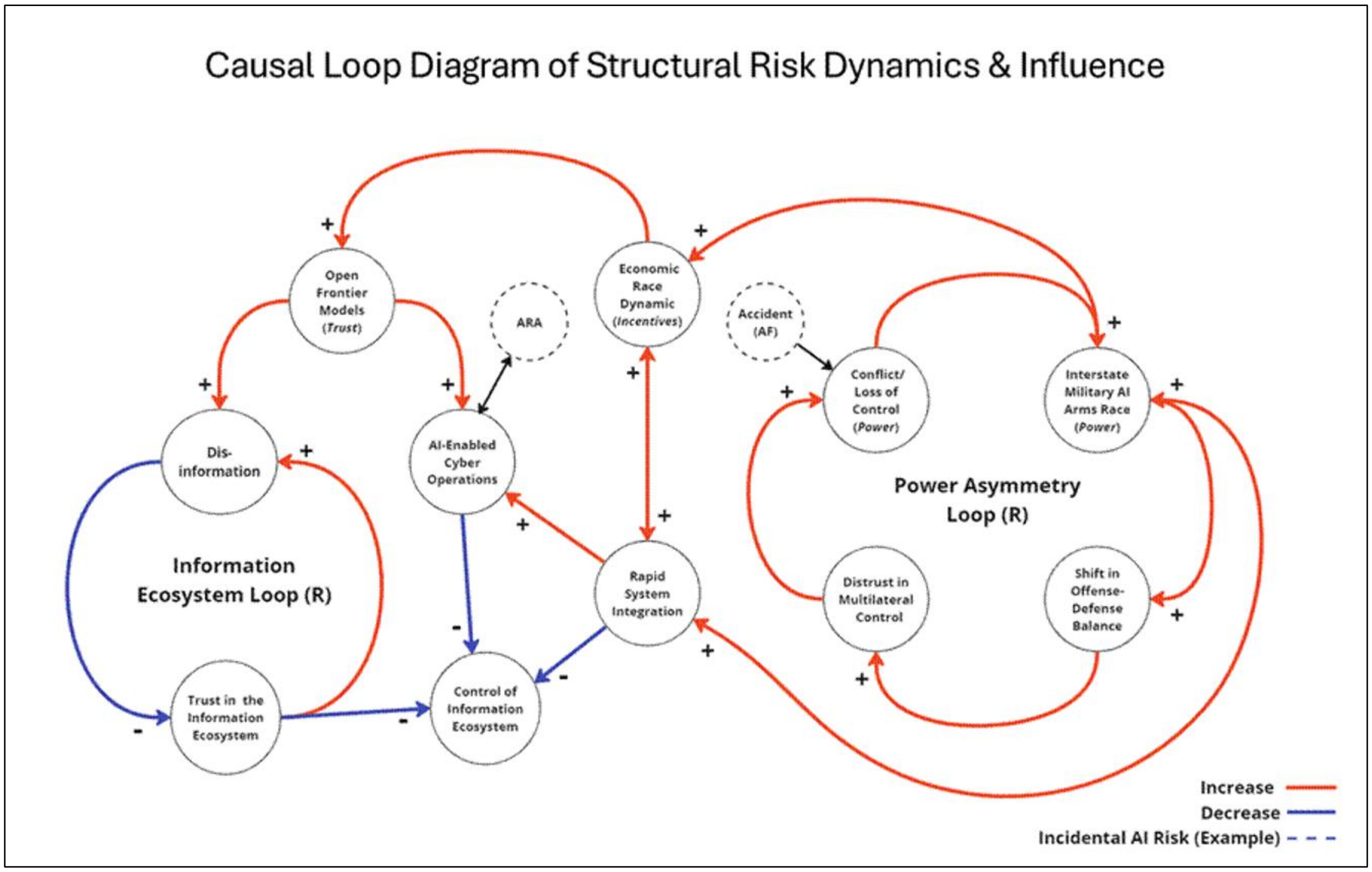

**Fig. 2.** An example causal loop diagram of complex causal pathways of structural drivers, accidents, and misuse. The plus symbols (red) indicate an increase (e.g., race dynamics increase the chance of conflict), and the negatives (blue) denote a decrease (e.g., disinformation decreases trust in the information environment). The figure displays two reinforcing feedback loops (the capital R denotes *reinforcing)* within the system: the information ecosystem loop and the power asymmetry loop (the dashed lines represent example proximate risks: ARA refers to autonomous replication and adaptation, and AF, alignment failure).

The persistent drive for power and control, especially in the national security context, incentivizes leaders to integrate AI across networked systems, with economic incentives accelerating these efforts. This is a virtuous circle in the best of times, but can spiral into vicious cycles in times of crisis (**Fig. 2**). Disruptive cyber intrusions into national systems have accelerated trends toward automating cyber defense in national networks, including using reinforcement learning (RL) agents (Nguyen & Reddi, 2019).[9] There have been equally ambitious proposals for intelligent next-generation networks (NGN), with the U.S. Army's Joint All-Domain Command and Control (JADC2) concept for networked operations and the U.S. Government's Resilient and Intelligent Next-Generation Systems (RINGS) program as two notable examples (Pomerleu, 2022; GCN, 2022). Rapid AI capability gains (potentially with ARA capabilities) coupled with expansive, intelligent networks could radically expand the digital surface for unexpected failures, cyberattacks, or misaligned systems beyond our control. These initiatives are driven, at least in part, by forces of power acquisition and control, distrust of adversaries, and underlying economic incentives. Thus, understanding and correctly interpreting the intersection of technical failure and structural forces is crucial for effectively managing and regulating advanced AI system integration.

Understanding these interlinked dynamics and their propagation across individual, organizational, state, and global levels is essential to identifying the structural fault lines that undergird AI deployment to design governance strategies that are both resilient and capable of adapting through stress.

[9] DARPA has pursued several large programs to leverage ML in defense of computer networks (DARPA, 2022). Reinforcement Learning (RL) based approaches have been proposed to monitor and evaluate CPSs for autonomous intrusion detection with multiagent RL-based simulations for defense against cyber-attacks (Nguyen, 2019).

# 5. Measuring & Mapping the Structural Risk Landscape

In light of these highly complex dynamics, measuring structural risk seems implausible or, at the very least, difficult and imprecise. Thus, this invariably involves qualitative judgments and proxy variables to help understand the trends and structural shifts underway. Nonetheless, in complex, highly uncertain problem spaces, imprecise, qualitative, and proxy values combined can be highly informative to help measure the degree of change and chart the best path forward for AI governance. These measures do not quantify any specific risk directly but rather the structural forces and influential variables that are known to be determinative. There are multiple approaches that could be valuable for measuring structural risk and sociotechnical change, such as causal loop modeling (similar to Fig. 2, see Crielaard, 2022), qualitative case study analysis (Cole, 2023), and a societal readiness approach (especially useful for technological resilience, see Lacroix & Thibault, 2024; Azemi, 2024) which follows the Technology-Readiness Level (TRL) paradigm developed by NASA (Hawkins, 2024). However, two stand out as useful for mapping the complexity and dynamics while informing policy. First, the development of a sociotechnical or structural risk index—with multiple variables indicative of structural change—could be helpful in evaluating the status of societal adaptation, the pace of change, and governance readiness at the state or global level. Second is strategic foresight—involving simulations, alternative futures, and wargaming—that evaluates interactions of structural forces, drivers, and agents that can help map the space of possible futures and understand how parameters interact to inform decision-making. Using structural risk index values (combined with other suitable parameters) to design simulations and scenario modeling could be crucial to understanding how these interactions and choices impact policies and change the risk landscape.

## *5.1 A Structural/Sociotechnical Risk Index*

A structural risk index could be invaluable for modeling, futures, and simulation design. These are measures of sociotechnical change that could be used in an analogous way to political risk indices. As an example, the index could be divided into four categories or dimensions: AI Diffusion (i.e., proliferation or propagation),[10] Integration, AI Policy, and Dependence. Much of this data is widely available from the OECD, Stanford's AI Index, the UN, and other sources.[11] Each of these dimensions has sub-categories or variables that could be measured through distinct quantitative indices, statistics, press reporting, or online activity

The index is intended as a tool for policymakers, strategic planners, intelligence or risk analysts tasked with anticipating and managing systemic risk from AI. This approach could be key to operationalizing the structural risk framework presented here (see Chapter 4) with the flexibility to integrate new, relevant parameters as they emerge. Much like political risk indices (Howell, 2016; Marchetti & Vitale, 2014) are used by governments (Bekaert et al., 2014), international organizations, and corporations (Hassan et al., 2014) to assess the stability of a region or regime this index would provide a standardized way to evaluate the sociotechnical volatility associated with AI adoption and integration. It could help states monitor emerging structural risks, compare governance readiness across sectors or nations, and prioritize investments in regulatory capacity, resilience planning, and international coordination. By providing a snapshot of AI-induced structural pressures, the index supports decision-makers in identifying early warning signs and taking preventive action before risks escalate or compound. **Table 3** details an example approach to mapping the dimensions, variables, and potential data sources (e.g., in the fourth column).

[10] In the context of technological propagation, diffusion refers to the process by which a new technology spreads across individuals, organizations, or societies over time; diffusion encompasses the stages by which technologies are adopted, integrated, and normalized across various sectors, influenced by social systems, with existing infrastructures (Zhang et al., 2023; Rogers, 2003).

[11] For the OECD's AI Policy Observatory, see here, Stanford's AI Global Vibrancy data sources here, United Nations Statistics here, or Transparency International's Corruption Perception Index here.

**AI Structural Risk Index**

| Dimension | Variable | Desc | Metrics |
|---|---|---|---|
| **Diffusion** | Acceleration | The pace of AI development, including growth in novel innovations | • Total increase in AI publications, partnerships, patents, new AI models, and investment<br>• Novel hardware and AI architecture development (percent change or number of innovations) |
| | Propagation | The spread of new AI technologies, paradigms, and adoption | • Pace of AI Knowledge flows and network growth (academic and industry)<br>• Pace of geographic distribution of skills migration and penetration |
| **Integration** | Commercial | The depth of integration globally, such as workforce penetration and talent concentration | • Change in AI industry, workforce, skills penetration, and hiring index<br>• Number of new AI startups and AI job listings |
| | Defense | AI investment in defense programs and the pace of new programs | • AI Investment in defense as a proportion of defense spending/GDP<br>• Increase in AI system integration in a military context |
| **Regulation** | Increase | The relative increase in AI policy programs for risk mitigation | • Increase in new policy proposals for AI governance and implementation<br>• International coordination, norms, and agreements |
| | Decrease | The degree countries are failing to enact or falling behind in AI mitigation policies | • Decline or stall in the number of policy programs, legal regimes, or failure of adoption<br>• Poor record of adhering to treaty obligations or non-signatories of safety regimes |
| **Dependence** | Individual | The rate of AI dependence across society, especially individual consumers | • Smartphone and social media addiction index; adoption of generative AI in the workplace<br>• Metrics on the degradation of human decision-making |
| | Organizational | The rate of AI adoption and job displacement in industry | • Job loss to automation without an accompanying increase in new jobs<br>• AI automation statistics and AI-related job loss, public and private |

**Table 3**. This table shows an example structural risk index. This table includes dimensions or key drivers, variables, descriptions of each, and potential metrics and data sources to include for measurement. This index measures the direction of predictive forces that could lead to ungovernable technological change.

The first dimension, AI diffusion, has two key variables—acceleration and propagation. Acceleration measures the overall pace and spread of AI systems throughout society; indeed, the dominant factor influencing the tension and dynamic interactions of AI and society is the speed of capability gains. The rapid pace of AI development stresses institutional flexibility and resilience, society's adaptive capabilities, and brings the potential for unintended escalations. In a similar vein, propagation measures the spread of systems across individuals, industries, and states. A key component of this is the open source ecosystem and whether it is restrained or allowed to spread the most highly capable systems. The second dimension is AI system integration with two key variables: commercial and defense. As with the acceleration of AI capabilities, the speed and breadth of autonomous system adoption and integration have a massive impact on how well organizations, institutions, and states absorb the shock of rapid change or how well adversaries react to these changes in a military context. Propagation also influences the availability of advanced models to individual actors, innocent developers, or criminals. The sub-categories commercial and defense evaluate these two complementary components that have the tendency to drive feedback loops, further accelerating integration. The variable defense integration includes a measure of the relative increase in reported cases of AI system integration into military activities (e.g., UAS used in Ukraine or Israel in Gaza), potentially derived from public reporting and data on budget acquisitions.

Beyond the pace and spread of AI technologies, the third dimension is AI policy. With the sub-categories of increase and decrease, this measure evaluates the preparedness and good behavior of organizations and state entities, which undergirds global collective security. Increase seeks to measure the change (relative and total) in beneficial AI policy proposals and their implementation, domestically and internationally; while proposals alone cannot ensure that safety and stability are trending in the right direction, their implementation can. Thus, this metric aims to standardize the relative increase in policy implementation across companies, governments, and societies. The second variable, Decrease, evaluates essentially the opposite, including companies and states unwilling to compromise on security, implement best practices and good governance, or worse, disengage. This, of course, adjoins one of the feedback mechanisms triggered by economic and geopolitical race dynamics. Further, the index evaluates adherence to important treaties or agreements, such as the Nuclear Non-Proliferation Treaty (NPT), and its history of failures in governance (e.g., Transparency International's Corruption Perception Index, or CPI).

The final dimension is dependence, with variables individual and organizational. This dimension aims to measure societal dependence on AI—including individuals, our collective decision-making capability, and organizations. At the level of the individual, there are a growing number of metrics and indexes that measure the social impact of technology, technology dependence, and AI's influence on decision making (Ahmad et al., 2023). The more societies depend on AI systems, the more likely there are other risks further down the causal chain (e.g., decision, truth, and value erosion). The second variable, organizational, looks specifically at AI penetration across the economy, such as skills penetration, AI demand, talent concentration, and different automation measures (see: OECD.ai or Stanford's AI Index). While this element is quite similar to integration, its focus is squarely on how businesses operate and their need for human labor. How these measures evolve over time could be suggestive of our ultimate control. For example, decreases in talent demand and available AI jobs, but sky-high talent concentration and automation would be indicative of increasing societal-scale dependence on AI and structural shifts away from human labor.

Multiple sources of data and new explanatory variables could increase the robustness of the index and more effectively measure the dimensions and the direction of sociotechnical change. More precise measures on military dependence on AI, trust measures across alliances, and social dependence would be vital. Operationally, these variables should be normalized to a fixed scale (e.g., 1-10 or 1-100) so that each measure can be evenly compared as is standard in risk management. Qualitative values, such as press reporting on defense integration or policy implementation, will require subjective judgment informed by subject-matter experts to transform these elements into workable metrics for the scale (e.g., the percent increase in reporting, policy proposals, or laws enacted) while the quantitative data can be easily transformed into a fixed scale for comparative analysis. Integrating generative AI into this framework (e.g., LLM iteration) for collecting and evaluating qualitative data could be invaluable, provided the researcher accounts for the potential for model errors and "hallucinations,"[12] integrating quality assurance checks for accuracy at each stage of analysis. From this, a total metric can be calculated for our current state of affairs at the global level or the relative stability of individual states or organizations. Given the uncertainty of societal-scale

[12] A model *hallucination* refers to an instance where an AI system, particularly a language model, generates output that is syntactically plausible but factually incorrect, fabricated, or unsupported by its training data or real-world knowledge (Ji et al., 2023)

risks from AI and trends impacting international security, developing workable metrics to track change will be increasingly important.

### *5.2 Simulating Structural Dynamics*

Exploratory futures, scenario modeling, and simulations are ideally suited for interrogating uncertain, computationally irreducible problem spaces. These approaches have been used in areas as diverse as climate change (Boettcher et al., 2016), environmental management (Malekmohammadi, 2017), national security (Tzezana, 2017; Johansen, 2018), policy (Bryan & Lampert, 2010), and AI risk (Gruetzemacher, 2019; Kilian et al., 2022) using a wide array of methods. Computational modeling and simulations are used widely in complex systems research to understand how interdependent variables interact and affect the broader system. In this way, scenarios and simulations are leveraged to understand how the interactions of structural dynamics impact AI risk while informing how different policies and actions could shift structures. Simulation models can reveal counterintuitive dynamics that other methods may miss. Monte Carlo simulations (Swendsen, 1987) for scenarios, cellular automata (CA) (Adiananda, 2016), and agent-based modeling (ABM) (Bonabeau, 2022) are three notable examples that could be useful for AI structural risk. For evaluating how individuals, companies, and governments change the risk balance, agent-based modeling (ABM) could be especially valuable. There is very little work using simulation for AI risk and governance, with notable exceptions. By setting each *agent*—whether individual, company, malicious actor, or nation state—and defining rules of interactions, the simulation can be iterated hundreds or thousands of times to uncover behavioral dynamics, counterintuitive interactions, and the impact of choices (Quang et al., 2018; Bonabeau, 2002).

Setting up useful and informative parameters for ABMs—and, importantly, choosing the right quantity and category of agents—is a challenging and important step for producing useful, generalizable insights. If there are too many agents and too much complexity, the model loses its utility, potentially introducing errors into the environment; the goal is not to perfectly model reality but to approximate high-level interactions to arrive at general principles. One method could be mapping structural drivers using proxy variables for incentives, trust, and power to different agents (e.g., individual, nation state, or regional entities such as the UN) for running simulations. Another approach could be to transfer variables and insights from the structural risk index to inform the setup of the model and define the initial conditions of the state space. Varying different rules to understand the acceleration and diffusion of technology, policy choices, implementation (such as increase and decrease), and level of dependence could be ideal for establishing a baseline for simulation. This could also ground the simulation in data from our current environment to play out plausible near-to-mid-term future paths. How these simulations evolve and adapt can be presented as scenarios for policy and planning. This approach could provide a grounded framework for understanding the complex structural dynamics and their impact on social stability and advanced AI risk.

Following the simulations and scenario generation, wargaming the decision environment and different contingencies could support companies, policymakers, and militaries seeking to understand the implications of rapid technological change. Games are useful for exploring complex systems—e.g., tradeoffs, tipping points, unique situations, and low probability risks—the kind of potential futures one expects with unprecedented change (Bartel, 2020; Crichton, 2000). Indeed, forecasting has critical limitations in this area, attempting to anticipate events that cannot be extrapolated from data. After formulating plausible futures and simulating the distribution, participants can systematically test these scenarios to interrogate decision points and policy changes to help develop more robust and resilient strategies. Understanding exponentiality and the unexpected is difficult for humans. Thus, simulating these situations and visualizing the range of outcomes—slow, moderate, fast, or super-exponential—could help foster intuition for a better-calibrated response.

With the increasing capabilities of LLM applications, participants should actively experiment with strategies for integrating LLMs into the analysis pipeline, from brainstorming drivers to scenario development to wargaming. In fact, researchers from AI developer Scale AI recently joined forces with the Center for Strategic and International Studies (CSIS) to help devise wargaming approaches using LLMs for the Futures Lab (Jensen et al., 2024). LLM methods are being explored across a range of research domains, including the social and physical sciences (Grossman et al., 2023), forecasting (Halawi et al., 2024), causal map generation (Kiciman et al., 2024), and scenario discovery. Indeed, through creative LLM prompting or specialized fine-tuning, there is a significant space for exploratory analysis. Thus, considering the substantive concerns and uncertainty with AI risks, using AI byproducts (such as LLMs) to understand structural risks could be invaluable in identifying and mitigating AI externalities (e.g., using AI to understand AI). Combining the proxy variables with simulations, scenarios, and gaming—enabled by the latest advancements in generative AI—could facilitate more thorough preparation for radically uncertain futures.

# 6. A Concentration on Structure for Policy

AI structural risks, accidents, and misuse can ultimately be clustered under the same umbrella; social forces drive the direction and magnitude of AI risk, while rapid emergent capabilities can unleash new risk categories and actors, shifting power dynamics. While the standard typology of accidents and misuse is easy to articulate, the underlying structural dynamics are fundamental and require careful attention in evaluating mitigation measures and policy opportunities. Indeed, shifts in system capabilities can shape the propensity for bad actors to misuse AI systems, increasing their overall effectiveness, while the larger forces of geopolitical competition can exacerbate both, driving escalatory feedback. Understanding these interdependencies is crucial to managing the uncertainty and system vulnerabilities in the face of rapid technological change. Moreover, changes to structural dynamics can be either the driver or impediment (given the right mitigations) to whether or how risks manifest. Effectively targeting the touchpoints, at the level of interactions, between advanced AI systems and the environment will ultimately help influence its impact on international security and stability. As advanced AI continues to integrate across social, political, and economic domains, understanding and addressing structure becomes paramount.

There are means to take structure into account when managing the safe development and deployment of advanced AI. First, it is crucial to raise awareness of the system dynamics underlying AI risks from corporate entities to military planners and the policy community. Indeed, with a strong understanding of the dynamics of AI structural risk, policymakers can design instruments that anticipate structural risks, dangerous feedback loops, and unintended consequences. While it is unlikely that specific effects could be understood sufficiently in advance, mapping out the plausible paths can help guide decision-making to prepare for the unknown. Moreover, government bodies and research institutes should double down on futures and foresight research so that decision-makers can understand the broad contours of structural risks and map these onto policy choices. The research approach proposed in this paper, developing a structural risk index and using this to implement simulations for scenario mapping, could go a long way in understanding the subtle touchpoints, possible unintended impacts, and areas for targeted interventions. This could be improved through tabletop exercises and even wargaming to understand the extent of possible dangers, test decisions and solutions, and iterate policy through games and simulations. This approach changes the framing from a search for perfect solutions and reactive mitigations to planning for resilience.

At the system deployment and integration level, leaders need a perspective of antifragility—developing frameworks capable of being resilient to unintended shocks, failure modes, and power asymmetries. Ensuring that high-risk systems of concern are integrated with multiple redundancies and backstops to avoid cascading failure modes will be critical as advanced AI is webbed through interconnected systems. The so-called "Flash Crash" of the New York Stock Exchange (NYSE) in 2010 (Treanor, 2015), where the market dropped over a thousand points in minutes (due in part to misinterpreted signals of algorithmic trading systems), is an important reminder of how isolated system failures can spread quickly, with spiraling feedback, causing significant, unanticipated damage. In a similar vein, significant research should be expended into developing stopgap measures (e.g., kill switches, limited lifespans) carefully designed to stop an automated agent system (e.g., ARA) in the event of an inadvertent release, attack, or unexpected adaptation. By integrating stop-gap measures down the chain of causation, there would be greater chances of isolating networks and infected systems. This *safety-by-design* approach is gaining ground in cybersecurity, and the AI safety community should consider a similar framework (Ziadeh, 2024). While advanced, novel capabilities remain limited today, leaders should not operate on the faith of select experts or wishful thinking, given the current pace of progress. Further, rather than a complete pause on large AI training runs (as called for in 2023, see: FLI, 2023), staging strategic pauses (e.g., at select mile markers) coupled with enhanced evaluation and red teaming throughout the development cycle could help ensure that safety is primary while also taking into account economic incentives and societies need for technological growth. The all-or-nothing cadence of uninterrupted AI progress is unsustainable.

In this context, the EU AI Act offers a relevant model of risk-based governance that could enrich the conceptualization of structural AI risks. By explicitly categorizing AI systems into risk tiers and regulating their deployment accordingly, the Act reflects an institutional approach that mirrors the precautionary principle—where potential for harm justifies preventative action even under uncertainty (European Parliament, 2024). This aligns with the safety-by-design ethos, which has long-standing roots in engineering and cybersecurity and emphasizes embedding safeguards during development rather than retrofitting them later (Stilgoe et al., 2013). Additionally, normative frameworks rooted in human rights and diversity—especially those emphasizing fairness, dignity, and non-discrimination—can guide the ethical integration of AI systems into society, particularly as these technologies increasingly influence decisions in sensitive domains (Latonero, 2018). These policy tools offer actionable insights for shaping resilient and inclusive governance structures that can respond to structural transformations catalyzed by advanced AI.

By integrating structural considerations into planning, policies can be designed that account for second-order and third-order effects—elements such as adversary reactions to AI integration, shifts in offense-defense balance, or societal dependence (e.g., facilitating unemployment)—so that institutions can be better prepared for disruptive change. This could include limits on the degree of integration of AI systems into safety-critical systems and prohibitions in areas such as nuclear command and control (C2) or early warning detection systems; indeed, in these areas, failures could be catastrophic or existential. As with all collective action, ensuring the agreement of diverse parties, especially in verification or enforcement, could be challenging. Cultivating international norms around these issues would be a crucial first step. To prevent unintended escalations from shifts to offense-defense balances, it would be critical for leaders to engage at the international level—as done during the Cold War with nuclear deterrence—to reinforce multilateral trust or agree to limits on societal-scale integration. An international body involving stakeholders at all levels, from developers to policymakers to government leaders, could be key to gaining trust and legitimacy for these important decisions (Gruetzemacher et al., 2023).

Of course, implementing robust policy frameworks—especially those that aim to mitigate structural risks—entails significant costs and long time horizons. International norm-building, regulatory development, and enforcement mechanisms are resource-intensive and often require sustained political will, bureaucratic coordination, and technical expertise (Brundage et al., 2020; Marchant & Wallach, 2015). Moreover, anticipatory governance approaches, while necessary, may be perceived as speculative or premature, making funding and prioritization difficult without clear and present risks (Guston, 2014). However, these upfront investments can dramatically reduce the downstream costs of failure, whether in the form of cascading infrastructure collapses, unintended escalation, or social disruption. To improve feasibility, policymakers should consider incremental approaches—such as milestone-based evaluations or modular safety requirements—that distribute the burden over time while allowing for adaptation as knowledge and capabilities evolve.

By taking structural dynamics into account with policy and planning, actors can more effectively address risks with complex interactions that tend to increase the degree of uncertainty while decreasing options. Planning through a *what-if* lens can help professionals respond to novel risks that will undoubtedly manifest from structural changes. Unifying structural issues, AI accidents, and misuse provide a useful conceptual framework for AI governance, forcing the examination of these interactions and indirect effects into the decision-making process. This necessitates a clear-eyed understanding of how ingrained human tendencies—such as the pursuit of power, greed, fear, and distrust—can direct AI development toward precarious states or trigger unintended crises. Moreover, this framework flips the policy process from reacting to emergencies to preparing for resilient long-term solutions. Thus, it is critical that we develop policy measures that integrate these considerations—testing and planning for unexpected feedback—to devise targeted solutions for stable and secure AI futures.

## *Acknowledgments*

The author would like to thank Toby Pilditch and Iyngkarran Kumar for their helpful comments and suggestions in earlier drafts of this paper. The Transformative Futures Institute sponsored this work.
Correspondence should be addressed to Kyle A Kilian at kyle.a.kilian@gmail.com.